\documentclass[onecolumn,aps,prb,superscriptaddress]{revtex4}
\usepackage{ae}
\usepackage[T1]{fontenc}
\usepackage[ansinew]{inputenc}
\usepackage{amsmath}
\usepackage{amssymb}
\usepackage{graphicx}
\usepackage{color}
\usepackage[colorlinks]{hyperref}

\begin{document}
\title{Selfoscillations of Suspended Carbon Nanotubes with a Deflection Sensitive Resistance under Voltage Bias}

\author{Anders Nordenfelt}
\email{anders.nordenfelt@physics.gu.se}
\affiliation{Department of Physics, University of Gothenburg, SE-412
96 G{\" o}teborg, Sweden}

\date{\today}

\begin{abstract}
We theoretically investigate the electro-mechanics of a Suspended Carbon Nanotube with a Deflection Sensitive Resistance subjected to a homogeneous Magnetic Field and a constant Voltage Bias. We show that, (with the exception of a singular case), for a sufficiently high magnetic field the time-independent state of charge transport through the nanotube becomes unstable to selfexcitations of the mechanical vibration accompanied by oscialltions in the voltage drop and current across the nanotube. 
\end{abstract}

\maketitle

In recent years, the  electronic and elastic properties of carbon-based nanostructures, such as graphene and carbon nanotubes (CNT), have been the subject of intense research. Due to their exceptional electronic and mechanical properties, being both light and extremely stiff, they hold promise for a number of future applications in nano-electromachanics (NEM). One of the recent results has been the realization of devices in which the electronic transport of suspended CNTs exhibit a strong sensitivity to mechanical bending. Such a NEM coupling was shown experimentally in the seminal work of V. Sazonova et al \cite{sazonova} and more recently in \cite{witkamp,lassagne}. It has also been studied theoretically in several papers \cite{yury, uchida}. In these devices, a semiconducting CNT is suspended over a gate electrode and the change in resistance is caused by a change of electronic doping determined by the distance between the CNT and the gate. In a previous paper \cite{anders} we considered their behaviour in the DC-current bias regime and we showed that, given any monotonic dependence of the resistance upon displacement, in the DC-bias regime there is (almost) always a critical magnetic field $H_c$ above which the static state of the nanotube becomes unstable to selfexcitations of the mechanical vibration accompanied by oscillations in the voltage drop across the nanotube. Rough estimates demonstrated that the critical magnetic field in this case could be as small as in the order of $10$mT. In this paper we generalize those earlier results to the Voltage Bias regime and we show that selfexcitations can be achieved by the means of a magnetic field provided that there is also some inductance in the system. Depending on the relationship between certain characteristic timescales of the system the critical electro-mechanical coupling parameter can be either positive or negative. The disposition of this paper more or less follows the structure of the previous work, and for a more thorough discussion on the basic phenomena we refer to \cite{anders}.\\
\begin{figure}
\begin{center}
\includegraphics[width = \columnwidth]{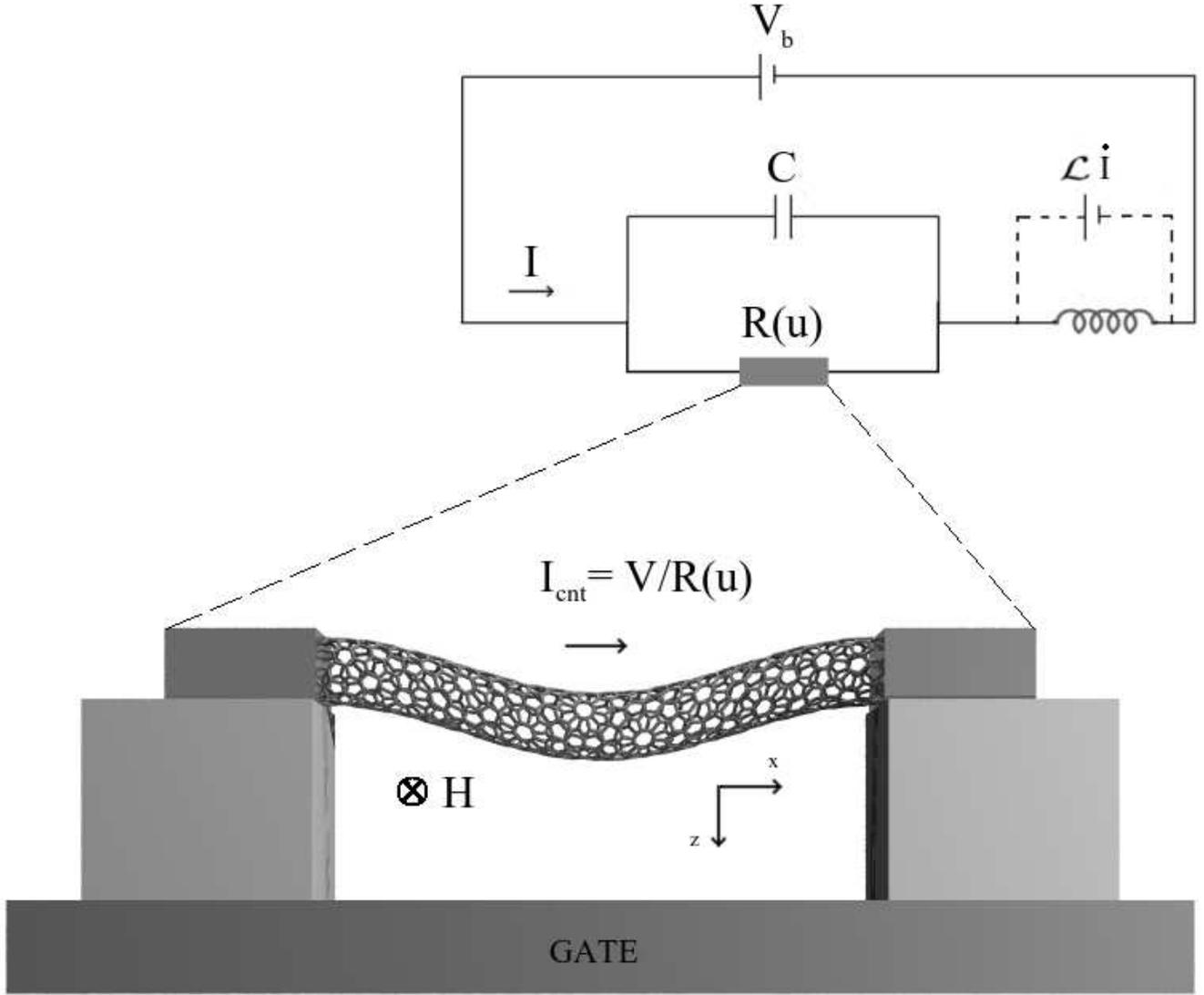}
\caption{The figure depicts a (semiconducting) carbon nanotube suspended over a gate electrode. The circuit is kept under a constant bias voltage and the nanotube is subject to a homogeneous magnetic field perpendicular to the direction of the current. Furthermore we assume that there is some capacitance and inductance in the system. A Lorentz force proportional to the curent through the nanotube and the magnetic field causes a deflection of the tube, which in turn changes the resistance. The inset on top shows the electric circuit.}\label{setup}
\end{center}
\end{figure}

The device is shown in figure ($\ref{setup}$). A (semiconducting) carbon nanotube is suspended over a gate electrode and subjected to a homogeneous magnetic field $H$ perpendicular to the direction of the current and parallel to the gate electrode. We assume that only the fundamental bending mode is excited to any appreciable degree and we make the Ansatz $z(t,x) = u(t)\varphi_{0}(x)$ for the mechanical deflection, where $\varphi_{0}(x)$ is the normalized profile of the fundamental mode \cite{LL}. This is motivated by the fact that the Lorentz force, which is the active feedback on the nanotube, is almost uniform across the tube and therefore have most effect on the fundamental mode. The conductivity of the nanotube is a function of the concentration of charge carriers which in turn is the product of the gate voltage $V_g$ and the mutual capacitance $C_g(u)$ between the nanotube and the gate electrode. The reason why the resitance $R(u)$ changes when the nanotube bends is because $C_g(u)$ depends on the distance between the nanotube and the gate. For our purposes, the important quantity to be drawn from the NEM coupling is the characteristic length defined by
\begin{equation}
\ell_R(u)= -\frac{R(u)}{R_u'(u)}.
\end{equation}  
The significance of this value will be shown below. Obviously, one can equally well consider other causes of the NEM coupling such as the effect of pure mechanical strain \cite{mechstrain1, mechstrain2} and the considerations below remain valid given any deflection sensitive resistance. The possibility of obtaining selfexcitations from other NEM-couplings will be discussed further at the end.\\ 

When the CNT is connected to an external voltage source $V_b$, a Lorentz force proportional to the current through the nanotube and the magnetic field causes a deflection of the CNT. As mentioned before, if the bending of the nanowire is on a scale much smaller than its length, the Lorentz force could be considered uniform and directed vertically at every point. If we assume that the mechanical motion is governed by the Euler-Bernoulli equation for an elastic beam, we arrive at the following set of equations for the time evolution of the voltage drop across the nanotube $V(t)$, the current $I(t)$ and the amplitude of the fundamental mode $u(t)$:
\begin{align}\label{main}
  &m\ddot{u} + \gamma \dot{u} +\kappa u = LHV/R(u) \\
  &C\dot{V} = I - I_{cnt}, \;\;\ I_{cnt} = V/R(u) \nonumber\\
  &\mathcal{L}\dot{I} = V_b - V \nonumber
\end{align}
Here $m$ and $L$ are the effective mass and length of the suspended part of the nanotube, $\kappa$ and $\gamma$ are the effective spring and damping constants, $C$ is the capacitance of the junction and $\mathcal{L}$ is the total inductance of the circuit. The system of equations (\ref{main}) has a time independent solution given by
\begin{align}\label{IV}
&u(t) = u_{0} = LHI_0/\kappa \\
&V(t) = V_{0} = V_b \nonumber \\
&I(t) = I_0 = V_b/R(u_0) \nonumber
\end{align}
A linear stability analysis of the time independent solution yields the following secular equation for the Lyapunov exponents $\lambda$:
\begin{equation}\label{secular}
P(\lambda,\beta) = (\lambda^2 + Q^{-1}\lambda + 1 - \beta)(\lambda^2 +\frac{\omega_{R}}{\omega_0}\lambda + \frac{\omega_{L}^2}{\omega_0^2}) + \beta \frac{\omega_{R}}{\omega_0}\lambda=0,
\end{equation}
where $\omega_{0}=\sqrt{\kappa/m}$ is the eigenfrequency of the fundamental mode, $\omega_{R}=1/(R(u_{0})C)$ is the RC-frequency, $\omega_{L} = \sqrt{1/(\mathcal{L}C)}$ is the LC-frequency and $Q=\sqrt{\kappa m}/\gamma$ is the quality factor \cite{zant}. The parameter $\beta$, referred to as the magneto-mechanical coupling parameter, is given by
\begin{equation}\label{beta}
\beta= \frac{LHV_b/R(u_0)}{\kappa\ell_R(u_0)}.
\end{equation}
Since $\beta$ depends on $u_0$ it is not necessarily directly proportional to the externally adjustable parameters $H$ and $V_b$. However, if the static deflection $u_0$ is small on the scale of the characteristic length, to a first approximation $\beta$ could be considered proportional to $H$. When some of the solutions to equation ($\ref{secular}$) have a positive real part, small deviations from the stationary solution ($\ref{IV}$) will grow exponentially during a certain time interval and eventually lead to a stationary state characterized by oscillations in all variables. By the use of for example the argument principle from complex analysis, one can show that two of the Liapunov exponents obtained from equation ($\ref{secular}$) have positive real part when
\begin{equation}\label{criterion2}
|\beta| > |\beta_c| ,\;\; \mathrm{Sgn}(\beta) = \mathrm{Sgn}(\beta_c)
\end{equation}
where the critical magneto-mechanical coupling parameter $\beta_c$ is given by
\begin{equation}\label{critical}
\beta_c = -\frac{1}{Q}\left( \frac{\omega_L^2 - \omega_0^2}{\omega_R\omega_0 + \omega_0^2/Q} + \frac{\omega_L^2/Q + \omega_R\omega_0}{\omega_L^2 - \omega_0^2} \right).
\end{equation}
We see that $\beta_c$ is negative when $\omega_L > \omega_0$ and positive when $\omega_L < \omega_0$, with a singularity obtaining for $\omega_L = \omega_0$. In the latter case there cannot be any instability regardless of the value of $\beta$. The physical meaning of the sign of $\beta$ is the following: If the coupling parameter is negative the carbon nanotube is pushed towards increasing resistance and if it is positive the nanotube is pushed towards decreasing resistance. Of particular interest is the low capacitance limit ($C \to 0$) for which equation ($\ref{critical}$) reduces to 
\begin{equation}\label{criterion1}
\beta_c = -\frac{1}{Q}\left(\frac{R}{\mathcal{L}\omega_0} + \frac{\mathcal{L}\omega_0}{R} + \frac{1}{Q}\right).
\end{equation}
By letting the capacitance go to zero we have in effect excluded the charge as a variable of the system and the relation between the voltage drop over the nanotube and the current is simply given by Ohm Law: $V = IR(u)$. As we can see, in this case the critical coupling parameter is always negative and if we assume a large quality factor it reaches a minimum absolute value approximately when $\omega_0 = R/\mathcal{L}$. This is in contrast to the Current Bias regime where the critical coupling parameter is always positive and reaches a minimum when $\omega_0 = 1/(RC)$ for high quality factors, see \cite{anders}. If no external capacitors or inductors are added to the circuit one can safely assume that $\omega_0 \ll 1/C, 1/\mathcal{L}$. It is thus evident that in the Voltage Bias regime, instability is more likely to occur at a low resistance while the opposite applies to the Current Bias regime. For a theoretical discussion on the self-inductance of a CNT we refer to \cite{wang}. Detailed calculations on the evolution of the instability are omitted here and we have to refer to computer simulations. It should be noted that, for a large characteristic length scale, the equations ($\ref{main}$) are only valid in a limited region of phase-space and in order to accurately descibe the time evolution outside this region one would have to include non-linear dynamical effects, especially in the mechanical degrees of freedom. \\

To conclude, in this paper we generalized our earlier results and showed that self-oscillations are possible also in the Voltage Bias regime provided there is some inductance in the circuit. One of the qualitative differences that showed up was the possibility of a negative sign in $\beta_c$. This circumstance might be of importance if we wish to utilize the effect of pure mechanical strain on the resistance. In this case the deflection dependence of the resistance is symmetric around the straight equilibrium configuration of the nanotube, hence, the sign of $\beta$ cannot be reversed by simply changing the direction of the magnetic field. Moreover, for metallic nanotubes the resistance is indeed likely to increase with bending \cite{mechstrain1}. \\

The author wishes to thank the Swedish Research Council (VR) for supporting this work. Discussions with L.Y. Gorelik and A.M. Kadygrobov are gratefully acknowledged.  

\pagebreak


\begin{thebibliography}{99}
\bibitem{sazonova} V. Sazonova, Y. Yaish et al. \emph{A tunable carbon nanotube electromechanical oscillator}. Nature, Vol. 431, (2004)
\bibitem{witkamp} B. Witkamp, M. Poot, H.S.J. van der Zant, \emph{Bending-mode vibration of a suspended nanotube resonator} Vol. 6, No. 12, 2904-2908 (2006)
\bibitem{lassagne} B. Lassagne, Y. Tarakanov, J. Kinaret, D. Garcia-Sanchez, and A. Bachtold \emph{Coupling Mechanics to Charge Transport in Carbon Nanotube Mechanical Resonators} Science 325, 1107-1110, (2009)
\bibitem{yury} Y. Tarakanov, J. Kinaret, \emph{A Carbon Nanotube Field Effect Transistor with a Suspended Nanotube Gate}, Nano Letters, Vol. 7 No. 8, 2291-2294 (2007)
\bibitem{uchida} K Uchida K, S Okada \emph{Electronic properties of a carbon nanotube in a field-effect transistor structure: A first-principles study}, PRL B, Vol 79, 8, 085402 (2009)   
\bibitem{anders} A. Nordenfelt, Y. Tarakanov, L. Gorelik et al. \emph{Magnetomotive Instability and Generation of Mechanical Vibrations in Suspended Semiconducting Carbon Nanotubes}, arXiv:1006.4477v1
\bibitem{LL} L.D.Landau and E.L.Lifshitz, \emph{Theory of Elasticity -3rd ed.}, Elsevier Ltd., 1986
\bibitem{mechstrain1} E.D. Minot, Y. Yaish et al. \emph{Tuning Carbon Nanotube Band Gaps with Strain}, PRL 90, 156401 (2003)
\bibitem{mechstrain2} J. Cao, Q. Wang, H. Dai \emph{Electromechanical Properties of Metallic, Quasimetallic and Semiconducting Carbon Nanotubes under Stretching}, PRL 90, 157601 (2003)
\bibitem{zant} A.K Huttel, G.A. Steele et al.,\emph{Carbon Nanotubes as Ultrahigh Quality Factor Mechanical Resonators}, Nano Letters, Vol. 9, No. 7, 2547-2552, (2009)
\bibitem{wang} B. Wang, R. Chu, J. Wang, H. Guo, \emph{First-principles calculation of chiral current and quantum self-inductance of carbon nanotubes}, Physical Review B, vol. 80, 23, 235430
\end{thebibliography}
\end{document}